\documentclass[12pt,preprint]{aastex}
\usepackage{psfig,natbib}
\shortauthors{Bower et al.}
\shorttitle{Galactic Center Radio Transient}
\begin{document}

\newcommand\degd{\ifmmode^{\circ}\!\!\!.\,\else$^{\circ}\!\!\!.\,$\fi}
\newcommand{\etal}{{\it et al.\ }}
\newcommand{\uv}{(u,v)}
\newcommand{\rdm}{{\rm\ rad\ m^{-2}}}
\newcommand{\msuny}{{\rm\ M_{\sun}\ y^{-1}}}
\newcommand{\mylesssim}{\stackrel{\scriptstyle <}{\scriptstyle \sim}}
\newcommand{\sci}{Science}


\title{A Radio Transient 0.1 pc from Sagittarius A*}

\author{Geoffrey C. Bower\altaffilmark{1}, Doug A. Roberts\altaffilmark{2,3}, Farhad Yusef-Zadeh\altaffilmark{3}, Donald C. Backer\altaffilmark{1}, W.D. Cotton\altaffilmark{4}, W. M. Goss\altaffilmark{5}, Cornelia C. Lang\altaffilmark{6}, Yoram Lithwick\altaffilmark{1}}

\altaffiltext{1}{Astronomy Department \& Radio Astronomy Laboratory,
University of California, Berkeley, CA 94720; gbower,dbacker,yoram@astro.berkeley.edu} 
\altaffiltext{2}{Adler Planetarium and Astronomy Museum, 1300 South Lake Shore Drive, Chicago, IL 60605}
\altaffiltext{3}{Department of Physics and Astronomy, Northwestern University, Evanston, IL 60208;
doug-roberts,zadeh@northwestern.edu}
\altaffiltext{4}{National Radio Astronomy Observatory, Charlottesville, VA 22903; bcotton@nrao.edu}
\altaffiltext{5}{National Radio Astronomy Observatory, P.O. Box O, Socorro, NM 87801; mgoss@nrao.edu}
\altaffiltext{6}{Department of Physics and Astronomy, University of Iowa, 203 Van Allen Hall, Iowa City, IA 52245; cornelia-lang@uiowa.edu}

\begin{abstract}
We report the discovery of a transient radio source 2.7 arcsec (0.1 pc
projected distance) South of  the Galactic Center
massive black hole, Sagittarius A*.  The source flared with a peak of at least 
80 mJy in March 2004.  The source was resolved by the Very Large
Array into two components with a separation of $\sim 0.7$ arcsec and characteristic sizes
of $\sim 0.2$ arcsec.  
The
two components of the source faded with a power-law index of $1.1 \pm 0.1$.
We detect an upper limit to the proper motion of the Eastern component
of $\sim 3 \times 10^3 {\rm\ km\ s^{-1}}$ relative to Sgr A*.  We detect a proper motion of 
$\sim 10^4 {\rm\ km\ s^{-1}}$ for the Western component relative
to Sgr A*.  The transient was also detected at X-ray wavelengths with the Chandra X-ray Observatory and the 
XMM-Newton telescope and given the designation CXOGC J174540.0-290031.  The X-ray source falls in 
between the two radio components.  The maximum luminosity of the X-ray source is $\sim 10^{36} 
{\rm\ erg\ s^{-1}}$, significantly sub-Eddington.  The radio jet flux density predicted by the 
X-ray/radio correlation for X-ray binaries is orders of magnitude less than the measured flux density.  
We conclude that the radio transient is the result of a bipolar jet originating in a single impulsive event
from the X-ray source and interacting with the dense interstellar medium of the Galactic Center.  

\end{abstract}


\section{Introduction}

The Galactic Center is home to a panoply of astrophysical objects, many of them interacting with each other.  At the center of this region is Sagittarius A*, a $3 \times 10^6\ M_\sun$ black hole radiating at $\sim 10^{-9}$ times the Eddington luminosity 
\citep{2001ARA&A..39..309M}.  Sgr A* currently has little effect on the surrounding medium aside from the effect of its mass.  Evidence suggests, however,
that this may not have always been the case.  The mass of Sgr A* and the Milky Way bulge velocity dispersion are consistent with 
the 
$M-\sigma$ correlation derived for more luminous active galactic nuclei (AGN), suggesting that Sgr A*, like other massive black holes and despite its 
current low power status, plays a central role in the construction of the Galaxy \citep{2002ApJ...574..740T}.  Moreover, the presence of strong reflected X-ray emission within 100 pc of Sgr A* suggests that activity in Sgr A* is episodic and that the very low luminosity of Sgr A* may not always be a constant feature \citep{2004A&A...425L..49R}.  Feedback related to star-formation and stellar death may play an integral role in regulating the activity of Sgr A*.  

This black hole is surrounded by a dense cluster of early-type stars on scales of 0.1 to a few pc \citep{2003ApJ...594..812G,2003ApJ...586L.127G}.  The winds from these stars may provide most of the 
mass that accretes onto Sgr A* \citep{1992ApJ...387L..25M,1999ApJ...517L.101Q}.  
The origin of these short-lived stars is a mystery \citep[e.g.,][]{1993ApJ...408..496M,2004ApJ...606L..21A}.  
Tidal forces are too strong at these radii, making {\em in situ} star-formation unlikely.  On the other hand, the time scale for these stars to fall from large radii 
through mass segregation or dynamical friction to their current location is much greater than the lifetime of the stars.   Co-spatial with these stars are complex streamers of ionized, atomic and molecular gas 
\citep{1996ApJ...459..627R,2005ApJ...620..287H}.
On a scale of 10 pc, the radio emission from the Galactic Center is dominated by Sgr A East, a young supernova remnant.  Interaction between Sgr A East and Sgr A* in the past may have triggered an episode of intense AGN activity 
\citep{2002ApJ...570..671M}.
On even larger scales, there is further evidence for significant star formation and for eventual stellar death in the form of supernova remnants \citep{2004AJ....128.1646N}.

Although the end-states of stellar evolution had been predicted to accumulate in the central parsec through the action of dynamical friction on objects 
formed in the central 10 parsecs \citep{1993ApJ...408..496M}, compact objects have not been   observed in these central regions until very
recently.
Deep integrations with the Chandra X-ray Observatory, however, have revealed an astonishing array of compact objects in the central 40 pc \citep{2003ApJ...589..225M}.  Further, 
the distribution of X-ray transients suggests the presence of an overdensity of compact objects in the central parsec of the Galaxy \citep{Muno05b...inprep}.

Observations of transient radio sources can provide important information about the compact object population near Sgr A*.  We report here on the discovery of a new radio transient only 0.1 pc in projected distance from Sgr A*.  We have
strong reasons to believe that the source is physically close to Sgr A*.
This source is closer by an order of magnitude than other radio  transients detected in the
Galactic Center \citep{1992Sci...255.1538Z,2002AJ....123.1497H,2005Natur.434...50H}.

The radio source was also discovered as an X-ray transient identified as CXOGC J174540.0-290031 \citep{Muno05...inprep,Porquet05...inprep}.  The X-ray properties of the object are:  (1), a factor of $\sim 100$ increase in apparent X-ray 
luminosity relative
to historical measurements at this position; (2), a significantly sub-Eddington luminosity $L_X = 10^{36} {\rm\ erg\ s^{-1}}$, the bulk of which is inferred from reflected X-ray emission along a ridge of Galactic Center dust and ionized gas; (3), significant absorption column depth that places the object at the Galactic Center or beyond; and (4), 
periodic X-ray emission, indicative of an eclipse by an edge-on accretion disk.  
Additionally, \citet{Muno05...inprep} report the absence of an infrared counterpart with an upper limit of $K < 16$, indicating that the mass-losing companion to this object is a late-type star. Thus, the likely classification of this object is as a low mass X-ray binary (LMXB).  

In Section 2, we present Very Large Array and Very Long Baseline Array
observations of the transient.  The radio and X-ray transient were
first detected as part of a multi-wavelength campaign observing
Sgr A*.  The transient was detected over a period of nearly 1 year with the VLA.  The transient was not detected with
the VLBA.  In Section 3, we present the light curve, spectrum, position and proper motion of the transient.  
In Section 4, we argue that the radio emission is due to the interaction of a jet originating from the LMXB with the dense medium of the Galactic Center.  We summarize in Section 5.

\section{Observations and Data Analysis}

\subsection{VLA Observations}
VLA observations were obtained over a wide range of dates and frequencies in
all four VLA configurations.
Imaging and analysis was particularly difficult because of the specific
nature of this source.  The transient is located very close to Sgr A*,
so that it is blended with this bright source in low resolution observations.
Additionally, the transient is co-located with significant diffuse
emission from the bar of Sgr A West.  Furthermore, there is significant
diffuse emission on much larger scales that is difficult to fully remove
with incomplete sampling of the $\uv$ plane.  As a consequence of 
all of these factors, variations in $\uv$ coverage lead to significant
systematic errors which are difficult to quantify.

Observations and results are summarized in Table 1.  We report the date of observation,
frequency, size of the synthesized beam for the image chosen for analysis, image rms noise,
flux density of Sgr A* ($S_{SgrA*}$), flux density of the Eastern and Western components of the transient
($S_{E}$ and $S_{W}$), and positions of the Eastern and Western components of the transient relative 
to Sgr A* in right ascension and declination ($\alpha_E$, $\delta_E$, $\alpha_W$, and $\delta_W$).

All observations
were obtained with the VLA in continuum or multi-channel continuum mode.  
With the exception of observations on 8, 17, 24 and 31 October 2004, all
observations consisted of tracks spanning $\sim 4$ hours, although not
all of the time was dedicated to the Sgr A*-field and there was
frequency multiplexing in some cases.  The four October
dates listed consisted of 1 hour snapshots and, therefore, had
substantially worse $\uv$-coverage.

All observations included observation of 3C 286 for absolute amplitude
calibration.  With the exception of observations in January 2004, the
data were amplitude and phase calibrated using the compact extragalactic
source J1744-3116.  For the January observations,  Sgr A* itself was used for
short-term amplitude and phase calibration.

Archival data were obtained to characterize the field before the appearance
of the transient.  Nine separate observations at 4.9 and 8.4 GHz spanning
10 weeks in June-August 1999 were combined to form single images at these frequency.
These data are discussed in more detail in \citet{2002ApJ...571..843B}.  To 
combine these images, we first measured the flux density of Sgr A* in each epoch,
subtracted that flux density from the visibility data, merged the visibility
data from each epoch, added the mean flux density of Sgr A* back into the
visibility data, and then imaged the combined visibility data.  This procedure is
necessary to provide high dynamic range images under the conditions of variable
flux from Sgr A*.  In addition,
43 GHz observations in C configuration from March 2000 were also analyzed.

Images were made of the field with a range of limits on the $\uv$ coverage.
Using the full $\uv$ coverage, the transient is difficult to identify for
the reasons given above.  
The transient is readily apparent in images with
minimum $\uv$ distance cutoffs of $50 k\lambda$.  
For higher resolution images, we also imposed a $\uv$ maximum cutoff of
$600 k\lambda$.
The transient is clearly detected as two components in the March 2004 data
(Figure~\ref{fig:qbanddiscovery}).  We fit three-component Gaussian models to the
$50 k\lambda$ images at the location of Sgr A* and the two components clearly detected in the
March 2004 data.  In many cases, the Western component of 
transient is not readily detected.  This is likely due to confusion and not
due to variability  in this source.  Where the imaging quality is highest, however,
we always detect both components (Figures~\ref{fig:qband},
\ref{fig:xband}, and \ref{fig:cband}).  For the archival data, we take a
conservative upper limit to the flux density of 5 times the rms noise
in the image.

The sources are resolved by the VLA at high resolution.  For the well-sampled
43 GHz data of March 2004 when the transient was at its peak flux density, we find  
sizes of $0.7\pm 0.1$ arcsec for the two components.  For the  6 cm
results from Nov 2004, we find sizes of $0.15 \pm 0.03$ arcsec for the two components.
For other epochs,
in which the sampling is not as complete, we do not attempt to fit sizes.

We measure the polarization in a few of the epochs in which the transient is well-sampled
and well-resolved from Sgr A*.  In particular, we measure the polarization in the combined
image of the four March 2004
epochs at 43 GHz and in the November 2004 epoch at 5 GHz.  Polarization is not detected in either
epoch with $1\sigma$ limits of 0.5\% and 0.2\%, respectively.

We plot our results in Figures~\ref{fig:lc}, \ref{fig:pos}, \ref{fig:timera}, and \ref{fig:timedec}.
Flux density and position shows greater scatter than expected based on the formal errors from our
fitting.  Due to the complexity of the imaging problem, we ascribe this to the fact that $\uv$ coverage
differences lead to systematic errors.  That is, true errors are larger than the formal errors.  We cannot
readily separate short-term variability in the flux density from additional systematic error.  The same
is true of errors in position and sizes.

\subsection{VLBA Observations}

We performed several VLBA observations to search for a compact counterpart to the transient 
(Table~\ref{tab:vlba}).  We also examined one archival data set.  Several of these observations
included additional telescopes including a single VLA antenna, the phased VLA, and the Green Bank
telescope.  Standard calibration techniques were used for these observations including {\em
a priori} amplitude calibration.  We imaged large fields of view to search for this transient 
and any other compact sources.  In the case of the 43 GHz observations, we imaged a $2 \times 2$ arcsec$^2$
field at the position of Sgr A* and a $4 \times 4$ arcsec$^2$ field at the position of the 
transient.  We combined observations in October and November 2004 to make a single image.

Sgr A* was detected with very high signal to noise ratio in all epochs, albeit heavily resolved due to interstellar scattering
\citep{2004Sci...304..704}.  No other source was detected.  
Peak and total flux densities for Sgr A* are low by factors of $\sim 3$ to 5 relative to 
contemporaneous VLA measurements.  This is a consequence of the fact that VLBA baselines
resolves out the heavily scatter-broaded image of Sgr A*.
To estimate upper limits for the flux density of any compact transient source components, we use
the measured rms noise, the ratio of the peak VLBA and total VLA fluxes for Sgr A*, and
the assumption that the transient source is located at the distance of Sgr A* (and, thus,
similarly scatter-broadened).  Due to
the lever-arm effect of scattering and the proximity of the Galactic Center hyperstrong
scattering screen to Sgr A* \citep{1998ApJ...505..715L}, 
more distant sources will be even more heavily resolved and
our limits will be even less constraining.
The $3\sigma$ upper limits for transient compact flux density are in the range 20 to 45 mJy.

\section{Results}

The images show clear detection of a transient radio source resolved into two
components.  The first radio detection of the transient occurred on 28 March 2004.
We can constrain the epoch of the initial outburst to be no earlier than
22 Jan 2004.  On that date, 22 GHz VLA observations detected no source at the
position of either transient to a $5\sigma$ upper limit of 3 mJy.  Archival
VLA data from 1999 at 4.9 and 8.4 GHz and from 2000 at 43 GHz confirm the absence of the
transient at levels of 4, 4, and 7 mJy, respectively.  These archival data represent close matches
in $\uv$ coverage to the observations of 2004 and demonstrate that the transient is not
the result of imaging artifacts.

There is a faint diffuse structure slightly North of the position of the transient apparent in
the archival 8.4 GHz image (Figure~\ref{fig:xband}).  This structure is likely unresolved emission from
Sgr A West but we cannot exclude the possibility that it is associated with previous activity in
the transient source.

The light curve of the Eastern component can be fit as a power-law decay with index $\alpha_{E}=-1.1 
\pm 0.1$, for $S\propto t^\alpha$.  We find $\alpha_{W}=-1.1 \pm 0.05$ for the Western
component.  If we extrapolate the light curves to 22 January 2004, we estimate 
the maximum possible flux density of the components to be $\la 400$ mJy.  

It is difficult to assess the spectrum of the sources from most of the data.  Since the components are
resolved by the VLA, any set of observations at a specific epoch over a range
of frequencies has a degeneracy with respect to the source size and spectrum.
Using the full resolution of the array, we see the source
become resolved at high frequencies.  Tapering the 
$\uv$ distribution to exclude very long baselines matches resolutions but
reduces sensitivity significantly.  Furthermore, we cannot determine with certainty
whether we are completely separating the transient emission from the background
emission from the Sgr A West bar.  We are able to estimate the spectral index ($S \propto \nu^\alpha$)
from the 
22 and 43 GHz observations of 10 February 2005.  These observations have sufficient resolution
to separate the transients and sufficient sensitivity to allow a reasonable match of
beam sizes.  We find spectral indices of East and West components to be $-0.7 \pm 0.1$ and
$-0.9 \pm 0.1$, respectively.

The mean angular offset of the two components from Sgr A* are  
$\Delta(\alpha_{E},\delta_{E})=(+0.258 \pm 0.016,-2.658 \pm 0.049)$ arcsec and 
$\Delta(\alpha_{W},\delta_{W})=(-0.304 \pm 0.030,-2.995 \pm 0.095)$ arcsec.
This gives a total separation between
the components of $0.655 \pm  0.113$ arcsec.  The absolute positions (epoch J2000) of the Eastern and Western transient components are
(17h45m40.266, -29d00'30''.758) and (17h45m40.305s, -29d00'31''.095), respectively, assuming an epoch 2005.2 position for Sgr A*
of (17h45m40.04s, -29d00'28''.100) 
\citep{1999ApJ...524..816R}.
The X-ray
transient point source position is (17h45m40.03s,-29d00'31.0'') with a 90\%-confidence error circle
of 0.3 arcsec \citep{Muno05...inprep}.
This position falls at the intermediate point between the two radio components
(Figure~\ref{fig:pos}).

We measure  the
proper motion of the Eastern component in right ascension and declination
as (-0.033,-0.101) $\pm (0.003,0.008)$ 
arcsec/yr.  For the Western component we measure (+0.055,-0.319) $\pm (0.009,0.011)$ arcsec/yr.
Errors here are estimated from the scatter in the positions after removal of the 
best fit.  
Formally, these constitute significant detections of the proper motion in both axes.
However, given the wide range of frequencies and the variety in $\uv$ coverages we are reluctant to
consider any of the proper motions
real with the exception of the proper motion of the Western component in declination.
A more accurate assessment of the error can be obtained as follows:  variations
within a few days give a characteristic error of $\sim 0.1$ arcsec per epoch; 
since systematic errors are dominant, the number of independent epochs is   reduced
to the number of observations conducted at different frequencies and in different
configurations; therefore, errors in proper motion are on the order of $0.05$ arcsec/yr.
We conclude that the Eastern component is stationary with an upper limit to the proper motion 
of $\sim 0.1$ arcsec/year while the proper motion of the Western component in declination has been
detected at a level of $-0.32 \pm 0.05$ arcsec/yr in a position angle $190 \pm 10$ degrees East of North.

\section{Discussion}

The radio emission is best explained as the interaction of a bi-polar jet
originating  from a low mass X-ray binary with the surrounding dense medium of the Galactic
Center.  The jet was probably launched prior to the first detection of the radio source and the first X-ray detection on 28 March 2004 but after our radio non-detection of 22 Jan 2004.  
A number of arguments and pieces of evidence support this picture.

\subsection{Location of the Transient in the Galactic Center}

The transient is almost certainly located close to Sgr A*.  First, X-ray spectra are best modeled with a large absorption column, $N_H \sim 6 \times 10^{22} {\rm\ cm^{-2}}$, consistent with the absorption column depth measured in the X-rays towards Sgr A* itself \citep{Muno05...inprep}.  This places the source at or beyond the Galactic Center.  Second, the density of stars and compact objects is known to peak at Sgr A* \citep{1987ARA&A..25..377G,2003ApJ...589..225M}.  It is most probable then that the source is located near the peak of the distribution rather than away from it.  

An extragalactic origin for the transient is unlikely based on the scale of the object.
The light travel time for a 1 arcsec source at 1 Mpc is 17 years.  The appearance of a new source of this size in a time scale of months can only 
appear with the effects of relativistic boosting.  For distances beyond 1 Mpc, the required Doppler boosting factor becomes rapidly unreasonable.

The absence of scattering produces a strong upper limit on the distance to this object.  A hyperstrong scattering region covers the central 15 arcmin of the galaxy \citep{1998ApJ...505..715L}.  The medium is estimated to lie at a distance of $133^{+200}_{-80}$ pc from Sgr A*.  Due to the proximity of the scattering medium to Sgr A*
(relative to the distance to the Sun), the efficiency of the scattering medium for Galactic Center sources is relatively small.  For sources further 
away from the Sun than Sgr A*, the scattering angle increases substantially.  For instance, at a frequency of 1.4 GHz, an extragalactic source will have a size of $\sim 60$ arcsec, which would be  unobservable since it will be blended with the diffuse structures of Sgr A.  The absence of strong scattering
places the transient close to the Galactic Center.  In particular, 
a size of 0.2 arcsec measured in November 2004 at 5 GHz provides a strong upper limit of a few hundred parsecs beyond Sgr A*.  

\subsection{Morphology and Proper Motion of the Radio Source}

The double structure of the radio source is strongly reminiscent of radio doubles observed in compact symmetric objects and other powerful extragalactic radio sources.  Moreover, the X-ray source is constrained to fall along the line connecting the two radio sources, suggesting that CXOGC J174540.0-290031 is at the position of the compact object and that the radio components are the working surfaces of a bipolar jet.  The light travel time for the 0.3 arcsec separation of each component from CXOGC J174540.0-290031 is 14 days.  This is consistent with an initial outburst that creates the radio components between 22 January and 28 March 2004.

We set upper limits to the proper motion of the Eastern component of $\sim 3 \times 10^3 {\rm\ km\ s^{-1}}$ and detect proper motion in the Western component at $\sim 1 \times 10^4 {\rm\ km\ s^{-1}}$.  These velocities are $\la 0.03c$, at least an order of magnitude less than the typical velocities in radio jets from X-ray binaries.  As we discuss in the next section, these velocities are characteristic of jets interacting with their medium, i.e. radio lobes.

\subsection{Large Radio Flux Density}

X-ray activity was measured at a level of $2 \times 10^{34} {\rm\ erg\ s^{-1}}$, within
a factor of two, with XMM in 28-31 March 2004 and 31 August-2 September  2004 \citep{Porquet05...inprep} and with Chandra in 5-7 July 2004 and
28 August 2004 \citep{Muno05...inprep,Porquet05...inprep}. The actual luminosity of the source may be significantly higher.  The reflected component of the X-ray emission that is observed suggests that the true X-ray luminosity is $L_X\sim 10^{36} {\rm\ erg\ s^{-1}}$.  
The absence of a detection by the RXTE All Sky Monitor places an upper limit to the X-ray flux at any time of $\sim 10^{36} {\rm\ erg\ s^{-1}}$ 
\citep{1996ApJ...469L..33L}. 
The steady decay of the radio light curve encompassing
these observations and the relative stability of the X-ray flux suggests that 
a single impulsive event prior to 28 March 2004 was responsible for the radio
emission.

The X-ray luminosity, the appearance of 8 hour eclipses in the X-ray light curve, and the absence of a bright infrared counterpart argue for a LMXB origin for this source.  It is not clear whether the compact object is a black hole or a neutron star \citep{Muno05...inprep,Porquet05...inprep}.  

A strong correlation between X-ray and radio luminosity for X-ray binaries exists during the outburst state \citep[e.g.,][]{2003A&A...400.1007C,2003MNRAS.344...60G,2004A&A...414..895F}.  These relations predict a radio flux density of $<< 1$ mJy for $L_X=10^{36} {\rm\ erg\ s^{-1}}$ and the case of a stellar mass black hole.  The radio-X-ray luminosity
correlation is a strong function of black hole mass.  Much of the discrepancy
in radio luminosity could be accounted by an increase in the mass of the compact object if it is increased to $10^3$ or $10^4 M_\sun$.  There is no
other evidence, however, that this object represents an intermediate mass black hole.  Additionally, astrometric measurements of Sgr A* have ruled
out the possibility of a $10^4 M_\sun$ companion at a separation of $\sim 10^{-2}$ to 1 pc \citep{2004ApJ...616..872R}.

These relations, however, are for the compact core radio emission of the jet.  The absence
of VLBA detection of compact emission indicates that the majority of the radio emission
comes from a source larger than $\sim 0.01$ arcsec.  Radio lobe emission has been modeled by \citet{2002A&A...388L..40H}.  Radio lobes are rarely detected in galactic X-ray binaries, because these jets are 
in relatively underdense media relative to their mean kinetic luminosities as compared to 
extragalactic radio jets.   The mean kinetic luminosity is the ratio of the kinetic
luminosity to the jet cross-section.  The ratio of mean kinetic luminosities between galactic and
extragalactic sources is proportional to the ratio of black hole masses, which $\sim 10^{5 --- 8}$.
Given a typical intergalactic medium density of $10^{-4} {\rm\ cm^{-3}}$,
\citet{2002A&A...388L..40H} infers interstellar medium densities 
for the environment of an X-ray binary to be much less than $1\ {\rm cm^{-3}}$.  

The much higher densities in the Galactic Center can lead to the production of radio lobes around X-ray binaries.  Ionized gas densities in Sgr A West are $\ga 10^3\ {\rm cm^{-3}}$ 
\citep[e.g.,][]{1996ApJ...459..627R},
while molecular gas densities are as high as $\sim 10^6\ {\rm cm^{-3}}$ 
\citep{2005ApJ...620..287H}.  Additionally, \citet{Muno05...inprep} shows that the transient
is sitting on the edge of a region of enhanced dust emission, indicative of a high density 
environment.
\citet{2002A&A...388L..40H} argues that jets in a medium of this density will quickly reach the 
Sedov phase.  Using the molecular gas density, $L_X=10^{36} {\rm\ erg\ s^{-1}}$, 
we estimate  that the radio lobe flux density is on the order of $1 (t/t_X)^{-0.9}$ Jy, 
where $t$ is the time after the initial outburst.  The time scale of the X-ray outburst $t_X$
is uncertain for this source although $t_X \sim 1$ day for other 
galactic black hole binaries \citep[e.g., GRS 1915+105][]{2004ARA&A..42..317F}.
This places the creation of the radio lobes within $\sim 10$ days of their discovery in the radio in late March 2004.  
The actual outburst from the compact object responsible for the creation of the jet that interacts with interstellar medium
must be $\sim 14 (v_j/c)^{-1}$ days prior to that interaction, where $v_j$ is the velocity of the jet.  
In this picture, the jet propagates through a low density medium before colliding with a
dense component of the medium.   While we favor
a jet model for its simplicity, we cannot rule out a spherical or quasi-spherical outburst in which the morphology of
the radio source is determined by the medium with which the ejecta interacts \citep{2004ApJ...615..432M}. 

The power law index of the decay time scale for the radio lobe predicted by this model is consistent with the observed index $-1.1 \pm 0.1$.  The steep spectral index measured for the transients are also consistent with the source being in  a state of optically thin cooling.
Additionally, the proper motion measured for the Western component is $\sim  10^4 {\rm\ km\ s^{-1}}$.  This sub-relativistic velocity and the lack of a measured proper motion for the Eastern component are consistent with the jet strongly interacting with the dense medium.

We estimate the total energy radiated from the radio lobe as the peak radio luminosity of $3\times 10^{31}$ ergs/sec integrated over $3\times 10^6$ s.  This is only a small fraction of the total jet energy estimated from the X-ray emission $\sim 10^{36} {\rm\ erg\ s^{-1}} \times 10^5\ s = 10^{41}$ erg.  

The two galactic sources for which persistent radio structures are detected are also located in the Galactic Center, GRS 1758-258 \citep{1998A&A...338L..95M} and 1E1740.7-2942 \citep{1999ARA&A..37..409M}.  These structures have a scale of $\sim 10$ pc, more than an order of magnitude larger than that associated with this transient.  The smaller scale of this transient is probably due to the higher density closer to Sgr A*.  Further, any larger scale, low surface brightness components will be blended with the extended structure of Sgr A West and Sgr A East.  

In another LMXB, we see a related phenomenon to that seen in this transient.
The variable galactic radio source Sco X-1 produces bipolar ejecta that move at relativistic velocities ($\beta \sim 0.5$) 
and  decay in flux density over the course
of hours  \citep{2001ApJ...558..283F}.  These are interpreted as the working surface of the jet with the surrounding medium, i.e. radio lobes. 
For the case of
a medium in which the jet pressure dominates the external medium pressure and both sides of the shock reach ultrarelativistic
conditions, the predicted hot spot velocity is $0.33c$, consistent with the observed 
velocity in Sco X-1 \citep{1979ApJ...232...34B}.  This solution does not represent well the Galactic Center transient, which has 
a characteristic velocity $\la 0.07c$, implying that the pressure of the interstellar medium is comparable to that of the jet.

\subsection{The Impact of the Transient on the Galactic Center}

The transient and 
Sgr A$^*$ are embedded within at least two populations of stellar systems. 
One is an evolved stellar population with an isotropic stellar light 
distribution whereas the other is a young population of stars (e.g. the 
IRS 16 cluster) showing characteristics similar to the well-known Arches 
cluster near the Galactic center. The winds from 
massive stars provide most of the mass that accretes onto Sgr A* 
\citep{1992ApJ...387L..25M,1999ApJ...517L.101Q}.
However, the nature of activity 
associated with any members of the evolved stellar cluster near Sgr A* is 
still unknown.
While the transient temporarily exhibits a significant luminosity, this 
luminosity is 
considerably less than that of a single O star ($\sim 10^{39} {\rm erg\ s^{-1}}$) .
This is true even in the case of an Eddington level outburst.

Nevertheless, the transient may have a significant local effect.
The spatial distribution of the ionized gas within a few arcseconds of 
SgrA* shows a 2$''$ hole, known as the mini-cavity, in one of the ionized 
streamers (the Bar) orbiting Sgr A$^*$ 
\citep{1990Natur.348...45Y}.
The kinematics of ionized gas associated with the mini-cavity shows 
a high velocity gradient near Sgr A* \citep{1996ApJ...459..627R}.
The origin of the mini-cavity is unknown.  Stellar sources 
or compact objects
with outflow activity may provide the physical conditions for the creation of this 
unusual substructure.

This transient is one of many X-ray transients that have been discovered with
Chandra in the central 17 arcmin.  Of the 7 transients discussed by \citet{Muno05b...inprep},
this transient is the closest to Sgr A* and it is the 
only transient has been detected at radio wavelengths.  
The other X-ray transients have peak X-ray emission comparable to that seen
for CXO 174540.0-290031, implying that compact radio emission for these sources
is unlikely to be detected at the mJy level, consistent with 
upper limits at 1.4 GHz.  There is also no evidence for long-lasting diffuse emission from
these sources although we cannot quantify the effects of confusion from Galactic diffuse 
emission.
The detection of the transient CXO 174540.0-290031 in the radio at $\sim 100$ mJy, 
therefore, is likely to be closely linked to the very dense gas environment 
in the vicinity of Sgr A*.

\section{Summary}

We have reported here the discovery of a transient radio source within 0.1 pc of Sgr A*.  
Based on X-ray and radio properties, the radio source appears to be the interaction of a jet with the dense interstellar medium of the Galactic Center leading to the production of radio lobes akin to those in the extragalactic radio sources.  
The results are consistent with
a total kinetic luminosity of the jet 
$\sim 10^{36} {\rm\ erg\ s^{-1}}$, consistent with the total energy released in the X-rays.
The presence of the source at this proximity to Sgr A* confirms the existence of the remnants of massive stars in the Galactic Center and supports the picture that stellar birth and death may have an effect on activity
in the central black hole.

\acknowledgements The National Radio Astronomy Observatory is a facility of the National Science Foundation operated under cooperative agreement by Associated Universities, Inc. 


\begin{deluxetable}{lrrrrrrrrrr}
\tabletypesize{\scriptsize}
\rotate
\tablecaption{VLA Observations of the Radio Transient}
\tablehead{\colhead{Date} & \colhead{$\nu$} & \colhead{Beam Size} & \colhead{RMS} & \colhead{$S_{SgrA*}$} & \colhead{$S_{E}$} & \colhead{$S_{W}$} & \colhead{$\Delta\alpha_{E}$} & \colhead{$\Delta\delta_{E}$} & \colhead{$\Delta\alpha_{W}$} & \colhead{$\Delta\delta_{W}$}  \\
			  & \colhead{(GHz)} & \colhead{(arcsec $\times$ arcsec, deg)} & \colhead{(mJy)} & \colhead{(mJy)} & \colhead{(mJy)} & \colhead{(mJy)} & \colhead{(arcsec)}& \colhead{(arcsec)}& \colhead{(arcsec)}& \colhead{(arcsec)}    }
\startdata
Jun-Aug 1999 & 4.9 & $0.9 \times 0.4$, 5 & 0.87      & $711 \pm 1$ & \nodata & \nodata & \nodata & \nodata & \nodata & \nodata \\
\nodata        & 8.4 & $0.5 \times 0.2$, 3 & 0.77     & $811 \pm 1$ & \nodata & \nodata & \nodata & \nodata & \nodata & \nodata \\
26 Mar 2000 & 43 & $1.1 \times 0.4$, 23 & 1.43 & $1091 \pm 3$ & \nodata & \nodata & \nodata & \nodata & \nodata & \nodata \\
22 Jan 2004 & 22 & $0.8 \times 0.5$, 73 & 0.63 & $914 \pm 1$ & \nodata &\nodata &\nodata &\nodata &\nodata &\nodata  \\
28 Mar 2004 & 43 & $1.1 \times 0.4$, 4 & 0.55 & $1803 \pm 1$ & $83 \pm 1$ & $48 \pm 1$ & $0.281 \pm 0.003$ & $-2.591 \pm 0.006$ & $-0.341 \pm 0.003$  & $-2.741 \pm 0.011$ \\
29 Mar 2004 & 43 & $1.0 \times 0.5$, 1 & 0.33 & $1797 \pm 1$ & $83 \pm 1$ & $41 \pm 1$ & $0.266 \pm 0.002$ & $-2.617 \pm 0.004$ & $-0.338 \pm 0.002$ & $-2.830 \pm 0.007$ \\
30 Mar 2004 & 43 & $1.1 \times 0.4$, 4 & 0.56 & $1801 \pm 1$ & $80 \pm 1$ & $48 \pm 1$ & $0.279 \pm 0.003$ & $-2.607 \pm 0.006$ & $-0.234 \pm 0.003$ & $-2.774 \pm 0.010$ \\
31 Mar 2004 & 43 & $1.0 \times 0.5$, 1 & 0.35 & $1796 \pm 1$ & $93 \pm 1$ & $45 \pm 1$ & $0.266 \pm 0.002$ & $-2.617 \pm 0.004$ & $-0.341 \pm 0.002$ & $-2.812 \pm 0.007$ \\
11 Jun 2004 & 43 & $1.2 \times 1.1$, 67 & 0.95 & $1593 \pm 1$ & $64 \pm 1$ & \nodata & $0.306 \pm 0.010$ & $-2.653 \pm 0.010$ & \nodata & \nodata \\
12 Jun 2004 & 43 & $1.2 \times 1.0$, 63 & 0.85 & $1297 \pm 1$ & $56 \pm 1$ & \nodata & $0.278 \pm 0.010$ & $-2.654 \pm 0.010$ & \nodata & \nodata \\
13 Jun 2004 & 43 & $1.2 \times 1.0$, 63 & 0.80 & $1417 \pm 1$ & $52 \pm 1$ & \nodata & $0.316 \pm 0.010$ & $-2.650 \pm 0.010$ & \nodata & \nodata \\
6 Jul 2004 & 43 & $2.5 \times 0.9$, 10 & 1.34 & $1333 \pm 1$ & $51 \pm 1$ & \nodata & $0.233 \pm 0.010$ & $-2.918 \pm 0.029$ & \nodata & \nodata \\
7 Jul 2004 & 43 & $2.5 \times 0.9$, 11 & 1.61 & $1718 \pm 2$ & $64 \pm 2$ & \nodata & $0.294 \pm 0.011$ & $-2.749 \pm 0.029$ & \nodata & \nodata \\
8 Jul 2004 & 43 & $2.5 \times 0.9$, 18 & 1.49 & $1432 \pm 2$ & $45 \pm 2$ & \nodata & $0.275 \pm 0.014$ & $-3.045 \pm 0.033$ & \nodata & \nodata \\
1 Sep 2004 & 43 & $0.27 \times 0.11$, -4 & 3.76 & $1244 \pm 7$ & $29 \pm 5$ & \nodata & $0.40 \pm 0.1$ & $-2.64 \pm 0.1$ & \nodata & \nodata\\
2 Sep 2004 & 43 & $0.14 \times 0.07$, 5 &  5.22  & $1333 \pm 6 $ & $15 \pm 5$ & \nodata & $0.35 \pm 0.1$ & $-2.80 \pm 0.1$ & \nodata & \nodata\ \\
3 Sep 2004 & 43 & $0.10 \times 0.05$, -5 & 2.92     & $1400 \pm 4$ & $15 \pm 5$ & \nodata & $0.22 \pm 0.1$ & $-3.15 \pm 0.1$ & \nodata & \nodata \\
1 Oct 2004 & 1.715 & $1.5 \times 0.5$, 10 & 3.76 & $466 \pm 2$ & $25 \pm 2$ & \nodata & $0.233 \pm 0.023$ & $-2.694 \pm 0.045$ & \nodata & \nodata \\
\nodata      & 1.665 & $1.6 \times 0.5$, 10 & 3.67 & $448 \pm 2$ & $24 \pm 2$ & \nodata & $0.172 \pm 0.023$ & $-2.713 \pm 0.046$ & \nodata & \nodata \\
\nodata      & 1.515 & $1.8 \times 0.6$, 10 & 1.19 & $302 \pm 1$ & $26 \pm 2$ & \nodata & $0.165 \pm 0.016$ & $-3.069 \pm 0.029$ & \nodata & \nodata \\
\nodata      & 1.435 & $1.7 \times 0.5$, 12 & 1.16 & $371 \pm 1$ & $30 \pm 1$ & \nodata & $0.169 \pm 0.013$ & $-2.881 \pm 0.030$ & \nodata & \nodata \\
4 Oct 2004 & 1.365 & $2.1 \times 0.8$, -3 & 4.16 & $301 \pm 4$ & $31 \pm 4$ & \nodata & $0.389 \pm 0.040$ & $-3.03 \pm 0.12$ & \nodata & \nodata \\
8 Oct 2004 & 1.475 & $2.6 \times 1.0$, 11 &  1.99    & $497 \pm 2$ & $38 \pm 2$ & \nodata & $0.212 \pm 0.027$ & $-2.902 \pm 0.070$ & \nodata & \nodata \\\\
\nodata	   & 8.4 & $0.55 \times 0.23$, 13   &  3.36    & $1608 \pm 2$ & $22 \pm 2$ & \nodata & $0.302 \pm 0.009$ & $-2.650 \pm 0.020$ & \nodata & \nodata \\
\nodata      & 22 & $0.7 \times 0.5$, 43    &  23.9 & $996 \pm 24$ & \nodata & \nodata & \nodata & \nodata \\     
\nodata      & 43 & $0.7 \times 0.5$, 42     & 20.0 & $705 \pm 21$ & \nodata & \nodata & \nodata & \nodata \\
17 Oct 2004 & 1.475 & $4.3 \times 0.8$, 29 &  2.40   & $513 \pm 2$ & $28 \pm 2$ & \nodata & $0.452 \pm 0.070$ & $-2.21 \pm 0.11$ & \nodata & \nodata \\
\nodata	   & 4.9 & $1.2 \times 0.4$, 30 &  1.74    & $914 \pm 2$ & $28 \pm 2$ & $11 \pm 2$ & $0.213 \pm 0.022$ & $-2.735 \pm 0.029$ & $-0.283 \pm 0.037$ & $-2.973 \pm 0.075$ \\
\nodata	   & 8.4 & $0.7 \times 0.2$, 30 &  1.78    & $1083 \pm 2$ & $20 \pm 2$ & $13 \pm 2$ & $0.270 \pm 0.018$ & $-2.705 \pm 0.025$ & $-0.287 \pm 0.035$ & $-3.001 \pm 0.067$ \\
\nodata	   & 22 & $0.7 \times 0.2$, 30  & 15.4      & $1528 \pm 14$ & \nodata & \nodata & \nodata & \nodata & \nodata & \nodata \\
24 Oct 2004 & 1.475 & $2.8 \times 1.0$, -15 & 1.95      & $565 \pm 2$ & $12 \pm 2$ & \nodata & $0.279 \pm 0.097$ & $-2.50 \pm 0.22$ & \nodata & \nodata \\
\nodata	   & 4.9 & $0.9 \times 0.4$, -12 &  2.21        & $886 \pm 2$ & $23 \pm 2$ & $3 \pm 2$ & $0.244 \pm 0.018$ & $-2.603 \pm 0.038$ & $-0.21 \pm 0.13$ & $-2.77 \pm 0.27$ \\
\nodata      & 8.4 & $0.5 \times 0.2$, -12 &  1.92    & $969 \pm 2$ & $11 \pm 2$ & $9 \pm 2$ & $0.186 \pm 0.019$ & $-2.617 \pm 0.040$ & $-0.251 \pm 0.022$ & $-3.041 \pm 0.046$ \\
31 Oct 2004      & 4.9 & $1.0 \times 0.3$, 22 &  1.78     & $742 \pm 2$ & $15 \pm 2$ & $9 \pm 2$ & $0.258 \pm 0.027$ & $-2.709 \pm 0.052$ & $-0.221 \pm 0.045$ & $-3.000 \pm 0.087$ \\
\nodata      & 8.4 & $0.7 \times 0.3$, 22 &   2.48    & $901 \pm 2$ & $8 \pm 2$ & $10\pm 2$ & $0.277 \pm 0.024$ & $-2.703 \pm 0.045$ & $-0.345 \pm 0.020$ & $-3.113 \pm 0.037$ \\
19 Nov 2004 & 4.9 & $0.6 \times 0.2$, 7 & 1.62 & $954 \pm 1$ & $16.6 \pm 0.2$ & $11.0 \pm 0.2$ & $0.249 \pm 0.001$ & $-2.685 \pm 0.002$ & $-0.290 \pm 0.001$ & $-3.040 \pm 0.002$ \\
10 Feb 2005 & 22 & $0.3 \times 0.2$, 57 & 0.21 & $1114 \pm 1$  & $5.7 \pm 0.3$ & $4.5 \pm 0.3$ & $0.251 \pm 0.008$ & $-2.655 \pm 0.008$ & $-0.224 \pm 0.012$ & $-3.000 \pm 0.011$ \\
\nodata     & 43 & $0.3 \times 0.2$, 4 & 0.26 & $1660 \pm 1$ & $3.1 \pm 0.3$ & $2.8 \pm 0.3$ & $0.262 \pm 0.006$ & $-2.696 \pm 0.006$ & $-0.266 \pm 0.005$ & $-3.043 \pm 0.005$ \\
\enddata
\end{deluxetable}

\begin{deluxetable}{llllrrrrr}
\tabletypesize{\scriptsize}
\rotate
\tablecaption{VLBA Observations of the Radio Transient \label{tab:vlba}}
\tablehead{\colhead{Date} & \colhead{$\nu$} & \colhead{Stations} & \colhead{Beam Size} & \colhead{Field of View\tablenotemark{a}} & \colhead{Sgr A* Peak Flux} &  \colhead{RMS} & \colhead{Transient Flux} \\
			  & \colhead{(GHz)} &                  & \colhead{(mas$\times$mas, deg)} & \colhead{(arcsec$\times$ arcsec)} &  \colhead{(mJy)} & \colhead{(mJy)} & \colhead{(mJy)}}
\startdata
29 Aug 1999 & 8.4 &  VLBA+Y1 & $14.4\times 5.6$, 13 & $3.3 \times 3.3$ & 158 &  2.2 & $<33$\\
16 May 2004 & 43  &VLBA+GBT & $2.1 \times 0.3$, 0 & $2.0 \times 2.0$, $4.1 \times 4.1$\tablenotemark{b} &166 & 0.7 & $<20$\\
23 Jun 2004 & 8.4 & VLBA+Y27 & $25.4 \times 9.7$, -13 & $22 \times 45$ & 188 & 2.9 & $<46$ \\
20 Oct 2004, 7 Nov 2004 & 8.4 & VLBA & $11.3 \times 5.8$, 1 & $8.2 \times 8.2$ & 78 & 0.9 & $<35$\\
\enddata
\tablenotetext{a}{The size of the field imaged, centered on Sgr A*, unless otherwise noted.}
\tablenotetext{b}{This field was centered on the position of the Eastern component of the transient.}
\end{deluxetable}

\plotone{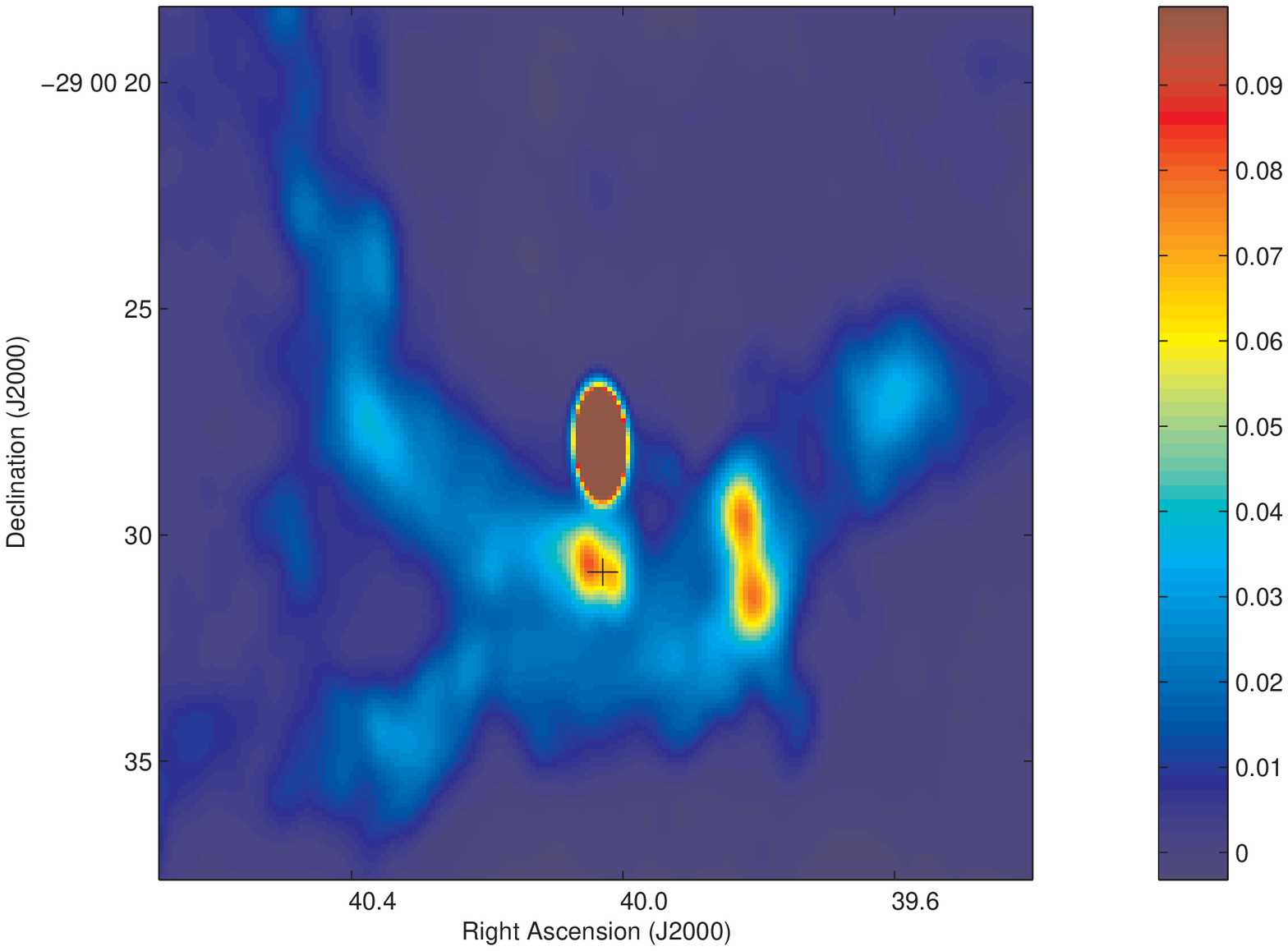}
\figcaption[]{Discovery image of the transient made with the VLA at 43 GHz from data obtained between
28 and 31 March 2004.  The transient is the double radio source due South of Sgr A*.  The mini-spiral of Sgr A
West is apparent along with IRS 13 and IRS 2 to the Southwest of Sgr A*.  The position of 
the X-ray transient CXO J174540.0-290031 has been marked with a cross, the size of which represents 90\%
confidence limits on its position.  The color scale has
been compressed with an artificial maximum of 100 mJy in order 
to bring out weak features relative to Sgr A*.  The beam size in this image
is $1.0 \times 0.5$ arcsec.  
\label{fig:qbanddiscovery}}

\mbox{\includegraphics[angle=-90,width=3in]{f2a.eps}
\includegraphics[angle=-90,width=3in]{f2b.eps}}
\figcaption[]{({\em Left}) C-configuration VLA image of Sgr A* at 43 GHz from 26 March 2000.  The transient is not present at a $5\sigma$ level
of 7 mJy.  ({\em Right}) C-configuration VLA image of Sgr A* and the transient at 43 GHz from 28-31 March 2004.  The transient is clearly present.
The position of the X-ray transient is marked with a cross as in Figure~\ref{fig:qbanddiscovery}.
Both images are made with baselines shorter than $50 k\lambda$ removed, in order to filter the large scale structure apparent in Figure
\ref{fig:qbanddiscovery}.  Contours are -10, -7, -5, 5, 7, 10, 14, 20, 28,  40, 56, 80, 112, 160, 224, and 320 mJy.  Beam sizes in the two images are $1.1 \times 0.4$ arcsec and 
$0.8 \times 0.5$ arcsec, respectively.
\label{fig:qband}}

\mbox{\includegraphics[angle=-90,width=3in]{f3a.eps}
\includegraphics[angle=-90,width=3in]{f3b.eps}}
\figcaption[]{({\em Left}) A-configuration VLA image of Sgr A* at 8.4 GHz from observations spanning June through August 1999.  
The transient is not present at a $5\sigma$ level
of 4 mJy.  ({\em Right}) A-configuration VLA image of Sgr A* and the transient at 8.4 GHz from 8-31 October 2004.  The transient is clearly present.
Both images are made with baselines shorter than $50k\lambda$ removed.  Symbols and contours are the same as in Figure~\ref{fig:qband}.  Beam sizes in the two images are $0.5 \times 0.2$ arcsec and $0.6 \times 0.2$ arcsec, respectively.
\label{fig:xband}}

\mbox{\includegraphics[angle=-90,width=3in]{f4a.eps}
\includegraphics[angle=-90,width=3in]{f4b.eps}}
\figcaption[]{({\em Left}) A-configuration VLA image of Sgr A* at 5.0 GHz from observations spanning June through August 1999.  
The transient is not present at a $5\sigma$ level
of 4 mJy.  ({\em Right}) A-configuration VLA+PT image of Sgr A* and the transient at 5.0 GHz from  19 November 2004.  The transient is clearly present.
Both images are made with baselines shorter than $100k\lambda$ removed.  This has the effect of removing most of the structure associated with IRS 13
and IRS 2 as well as Sgr A West.  Symbols and contours are the same as in Figure~\ref{fig:qband}.  Beam sizes in the two images are $0.9 \times 0.4$ arcsec and $0.6 \times 0.2$ arcsec, respectively.
\label{fig:cband}}

\plotone{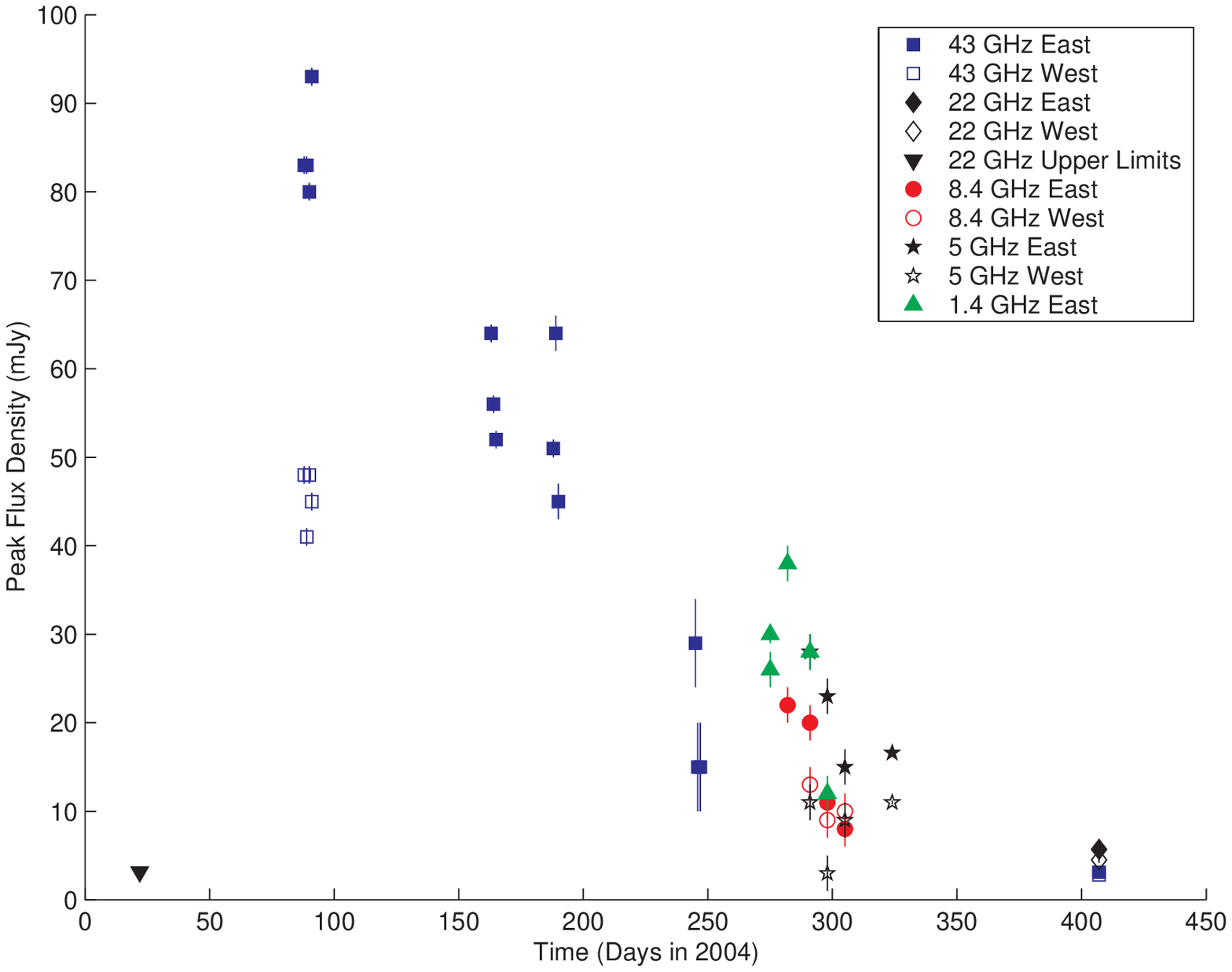}
\figcaption{Flux density of the transient components as a function of time for frequencies between 1.4 and 43 GHz.  Filled symbols 
represent the Eastern component while open symbols represent the Western component.  See the legend for frequency identification of symbols.
\label{fig:lc}}

\plotone{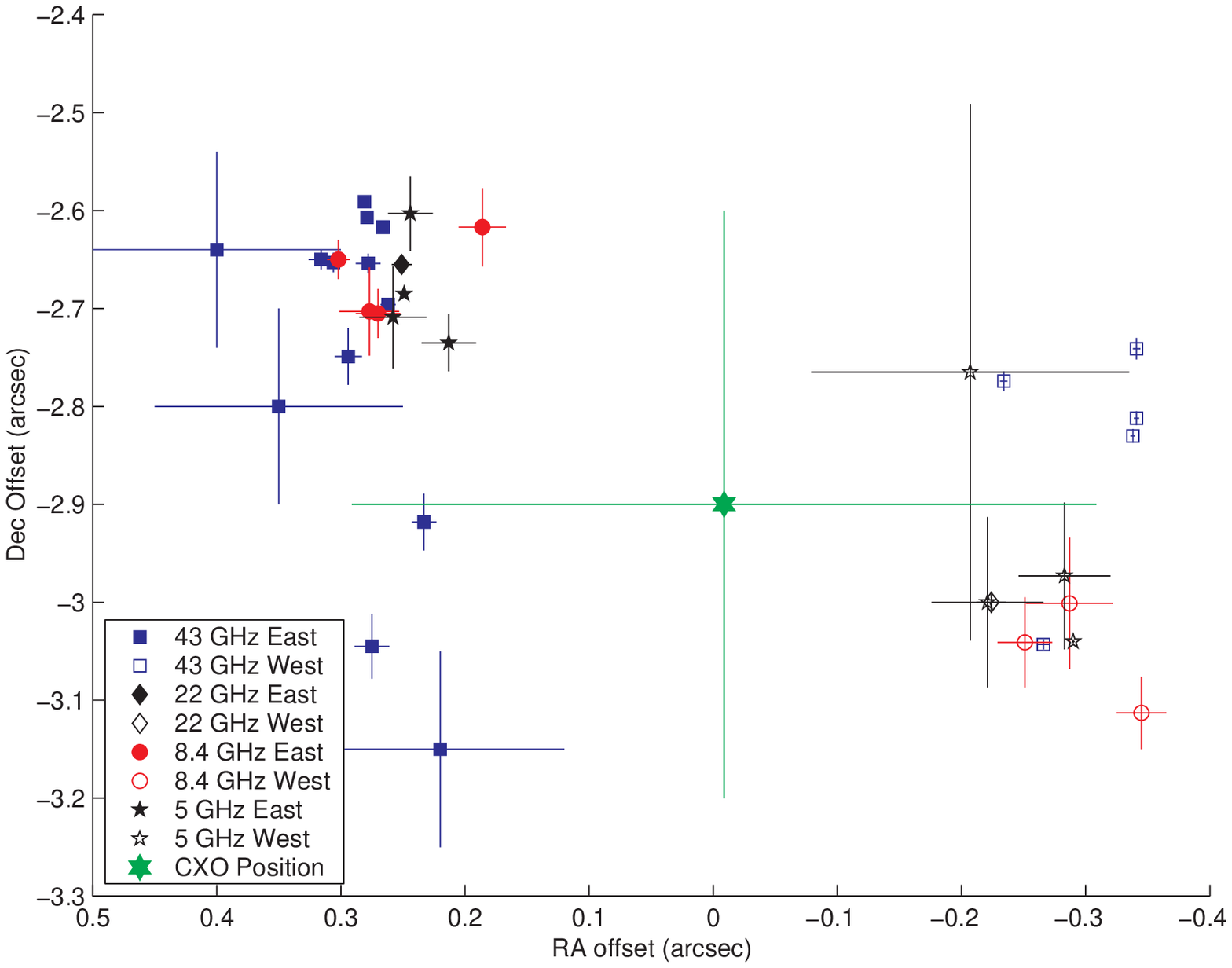}
\figcaption{Positions of Eastern and Western components of the transient 
relative to Sgr A*.  Symbols are as in Figure~\ref{fig:lc} with the addition of the
position of the X-ray source identified with Chandra.
\label{fig:pos}}

\plotone{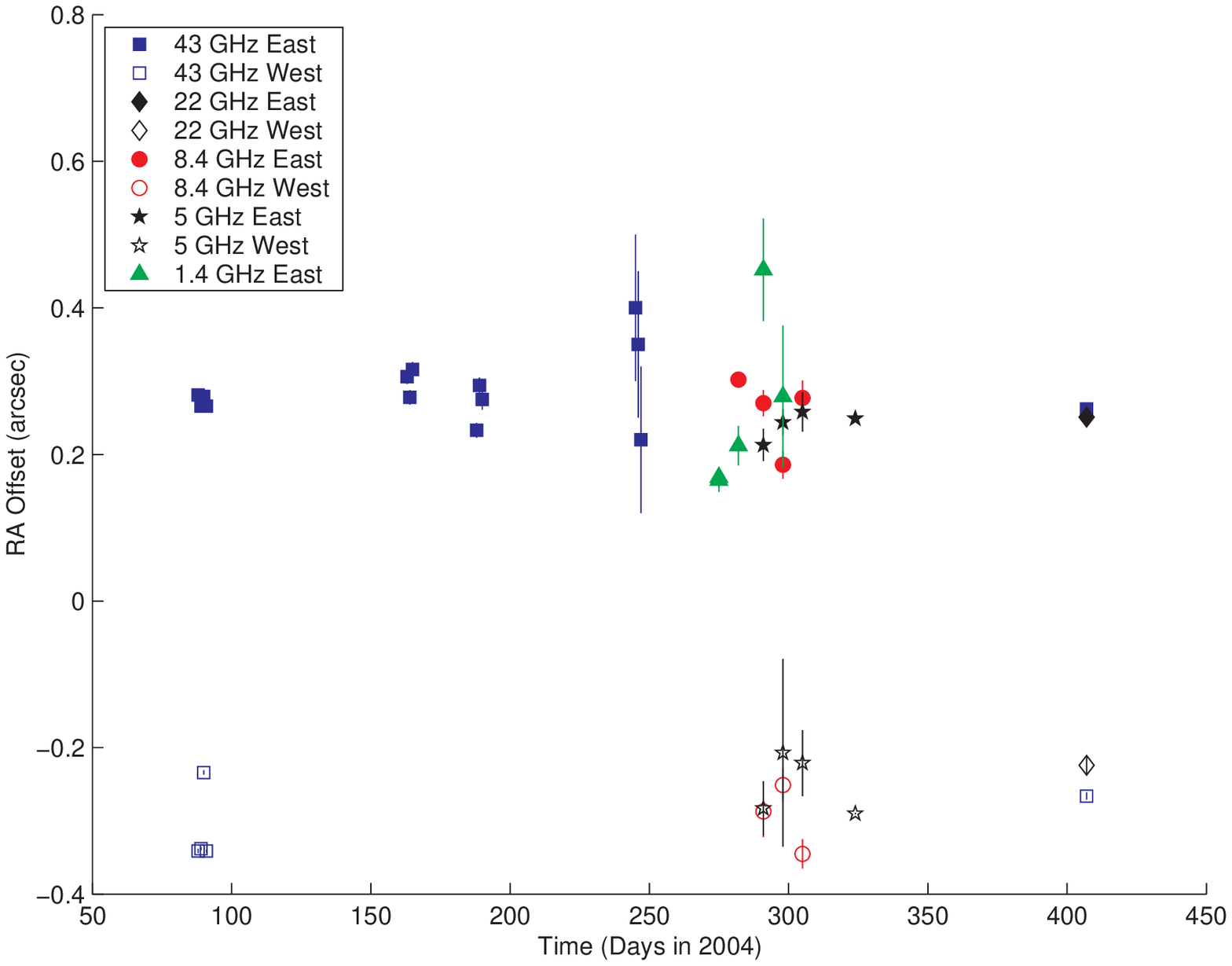}
\figcaption{Right ascension of Eastern and Western components relative to Sgr A* as
a function of time.  Symbols are as in Figure~\ref{fig:lc}.
\label{fig:timera}}

\plotone{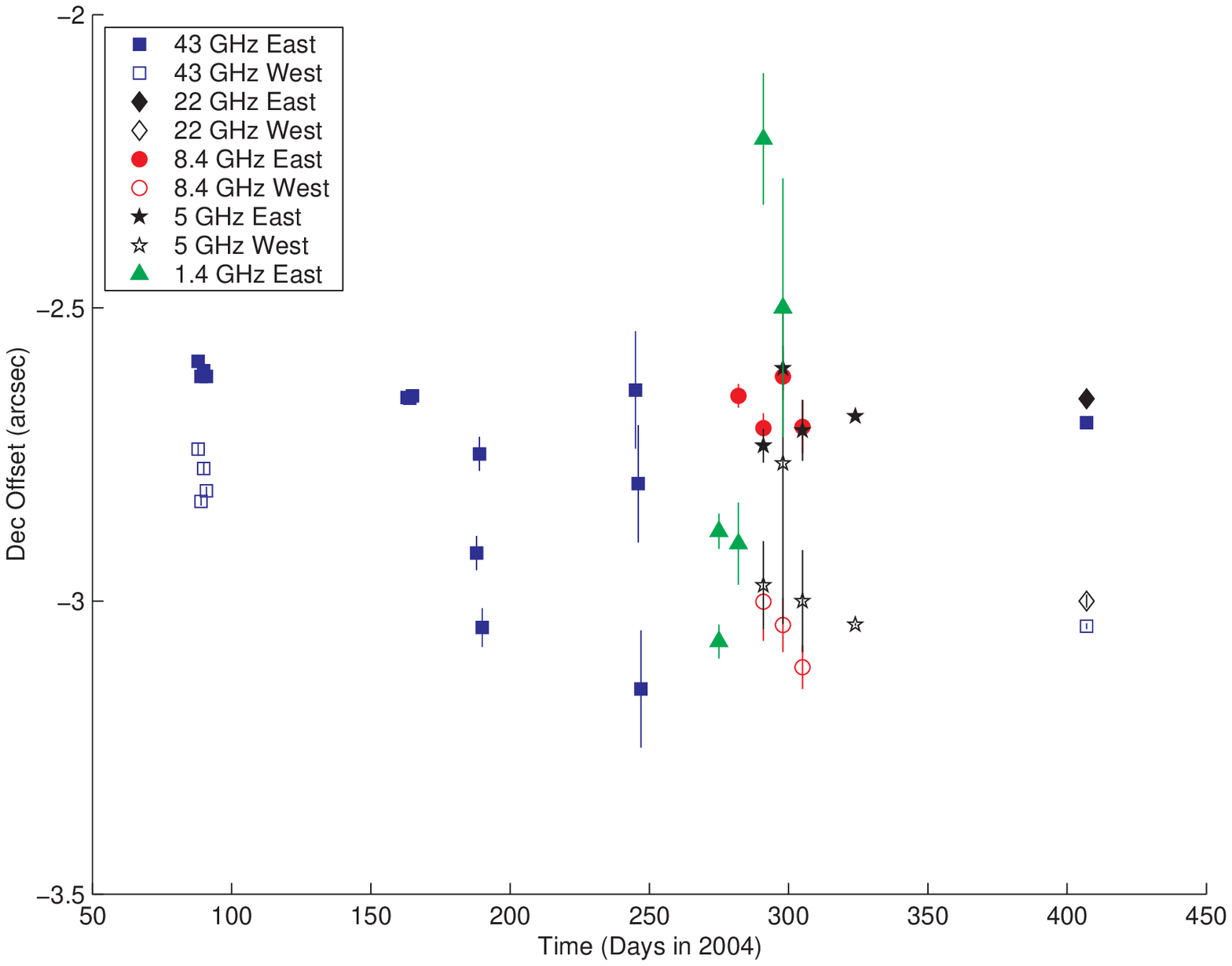}
\figcaption{Declination of Eastern and Western components relative to Sgr A* as
a function of time.  Symbols are as in Figure~\ref{fig:lc}.
\label{fig:timedec}}

\end{document}